\documentclass[11pt, a4paper]{amsart}

\usepackage{a4wide}
\usepackage{graphicx,enumitem}
\usepackage{amsmath,amssymb}
\usepackage[english]{babel}
\usepackage{subfigure}
\usepackage{units}
\usepackage{diagbox}
\usepackage{fontenc}
\usepackage{xcolor}
\usepackage{caption}
\usepackage{ulem}
\usepackage{cancel}
\usepackage{tikz}
\usepackage{amssymb}
\DeclareSymbolFontAlphabet{\amsmathbb}{AMSb}%
\usepackage{mathbbol}
\usepackage{latexsym}
\usepackage{xspace,bm}
\usetikzlibrary{positioning}
\usetikzlibrary{arrows,decorations.markings}
\usetikzlibrary {arrows.meta,bending}
\usetikzlibrary{external}
\usepackage[color=yellow]{todonotes}

\usepackage[pdftex,
bookmarksopen,
colorlinks,
linkcolor=black,
urlcolor=black,
citecolor=black
]{hyperref}
\hypersetup{pdfauthor={S.R. Ephrati, E. Jansson, K. Modin}}

\usepackage[noabbrev]{cleveref}

\usepackage{dsfont}

\usepackage{cases}

\usepackage{mathtools}

\newcommand{\R}{\amsmathbb{R}}

\newcommand{\N}{\amsmathbb{N}}

\newcommand{\IS}{\amsmathbb{S}}

\newcommand{\cE}{\mathcal{E}}

\DeclareMathOperator{\trace}{Tr}

\newcommand{\dd}{\mathrm{d}}

\newtheorem{lemma}{Lemma}[section]

\theoremstyle{remark}
\newtheorem{remark}[lemma]{Remark}

\theoremstyle{definition}

\newcommand{\ab}{$\alpha$-$\beta$}
\newcommand{\al}{$\alpha$}
\newcommand{\be}{$\beta$}

\usepackage{color}
\definecolor{darkgreen}{rgb}{0,.6,0}

\title{On spectral scaling laws for averaged turbulence on the sphere
}
\author{Sagy Ephrati \and Erik Jansson \and Klas Modin}
\date{\today}

\begin{document}

\maketitle
\begin{abstract}
Spectral analysis for a class of Lagrangian-averaged Navier--Stokes (LANS) equations on the sphere is carried out. The equations arise from the Navier--Stokes equations by applying a Helmholtz filter of width \al~to the advecting velocity \be~times. Power laws for the energy spectrum are derived and indicate a \be-dependent scaling at wave numbers $l$ with $\alpha l\gg 1$. The energy and enstrophy transfer rates distinctly depend on the averaging, allowing control over the energy flux and the enstrophy flux separately through the choice of averaging operator. A necessary condition on the averaging operator is derived for the existence of the inverse cascade in two-dimensional turbulence. Numerical experiments with a structure-preserving integrator confirm the expected energy spectrum scalings and the robustness of the double cascade under choices of the averaging operator.
\end{abstract}

\section{Introduction}

The distribution of energy over a vast range of scales of motion is a characteristic feature of turbulent flows. 
Consequently, fully scale-resolving numerical simulations of turbulence are computationally unfeasible. 
This has prompted the development of simulation strategies that deal with modified systems of partial differential equations with reduced dynamical complexity. 
Among these is a geometric regularization principle for ideal fluids as proposed by Holm, Marsden, and Ratiu~\cite{holm1998euler}, referred to as \al-modeling. 
It alters the nonlinear advection terms and thereby gives rise to regularized Euler equations. 
Adding a viscous term then leads to the Navier--Stokes-\al~(NS-\al, also referred to as Lagrangian-averaged NS-\al~or LANS-\al) model, in which the energy content of small scales is suppressed \cite{lunasin2007study}.

In this work, we study the two-dimensional NS-\al~model on the sphere and extend the model to a larger class of averaging operators, which we refer to as the \ab-Navier--Stokes (\ab-NS) equations.
More specifically, the contributions are the following.

\begin{enumerate}
    \item  We perform spectral analysis for averaged turbulence on the sphere to obtain scaling laws for the double cascade, using the geometric stream function-vorticity formulation.
    The analysis is carried out for the \ab-NS equations, which encompass both the standard NS equations and the NS-\al~model.
    By adhering to the stream function-vorticity formulation, we exploit geometric properties of the two-dimensional Euler equations and simplify previous work on the scaling of the Navier--Stokes equations \cite{lindborg2022two}.
    Specifically, without averaging we recover the double cascade of two-dimensional turbulence.
    When averaging, a steep energy scaling is found at high wave numbers, where the onset to this scaling is determined by \al~and the scaling rate is determined by the averaging strength \be.
    
    \item A necessary condition is derived for the inverse energy cascade to exist in averaged turbulence, which imposes certain conditions on the averaging operator.
    This ensures that the analysis is valid for any type of operator relating the vorticity and stream function that decays sufficiently fast for high wave numbers. 

    \item The energy and enstrophy flux rates are derived and are shown to distinctly depend on the chosen averaging operator. 
    The disparity between these flux rates permits control over the fluxes through an appropriate choice of averaging operator.

    \item The derived energy scaling laws are supported by numerical results carried out with a structure-preserving integrator, providing robust numerical support of the theoretically derived double cascade in averaged turbulence.
\end{enumerate}

The averaged Navier--Stokes equations are a model for the dynamics of the large-scale flow in a turbulent fluid. Here, the spatial domain is the two-dimensional sphere $\IS^2$ and the governing equations are given by
\begin{equation}
    \label{eq:averagedNS_momentum}
    \dot m + \nabla_u m + \nabla_m^T u = \nu (\Delta m + 2m) -\gamma u + f,
\end{equation}
where $\dot m = \frac{\partial m}{\partial t}$ denotes time derivative, $\nabla$ and $\nabla^T$ denote the covariant derivative and its transpose under the $L^2$-metric, $u$ denotes the averaged velocity field of the fluid, and $m = (1-\alpha^2 \Delta)^\beta u$ is the velocity, or momentum. 
The parameter $\nu \in \R^+$ determines the viscosity of the fluid, $\gamma \in \R^+$ is the friction parameter, and $\alpha$ is a length scale associated with the smoothing effect of the Helmholtz operator. 
The parameter $\beta \in \R^+$ determines the strength of the averaging, or, for integer values, the number of times the Helmholtz filter is applied. The addition of \be~marks an extension of the averaging operator when compared to preceding studies of the \al-model. 
We refer to \cref{eq:averagedNS_momentum} as the $\alpha$-$\beta$ averaged Navier--Stokes equations.

\begin{figure}[h!]
    \centering
    \includegraphics[width=0.98\linewidth]{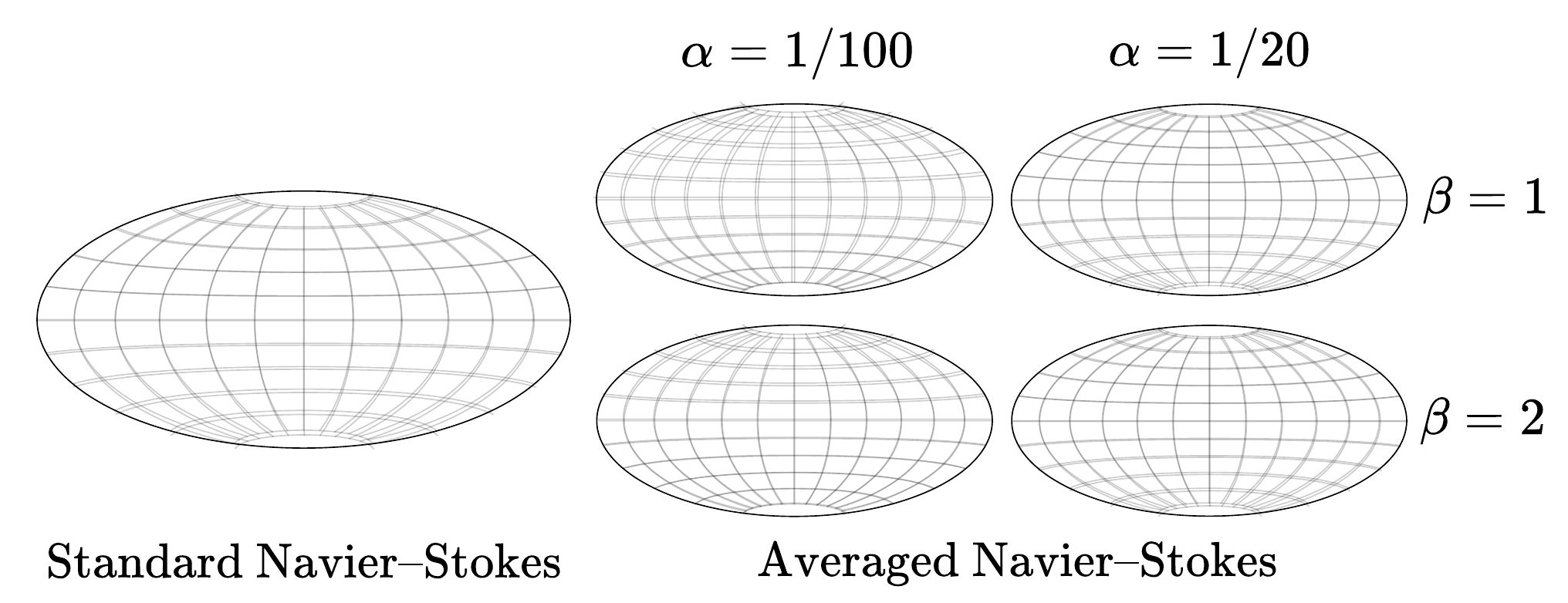}
    \caption{Demonstration of the smoothing effect on the vorticity.
    Left: Vorticity snapshot representative of the Navier--Stokes equations.
    Right: Smoothed fields after applying the \ab-filter to the vorticity field on the left.
    }
    \label{fig:vorticity_eyecandy}
\end{figure}

An equivalent formulation of \cref{eq:averagedNS_momentum} is the stream function-vorticity formulation. 
We distinguish between the velocity $u$ and the momentum $m$ via the relations $\nabla^\perp\cdot u=\omega$ and $\nabla^\perp\psi = m$, where $\nabla^\perp$ is the skew-gradient on the sphere, obtained by rotating the gradient by $\pi/2$ in the tangent plane of the sphere. 
The scalar functions $\omega$ and $\psi$ are referred to as the vorticity and the stream function, respectively. 
The formulation via $\omega$ and $\psi$ avoids having to deal with the covariant derivative terms in \eqref{eq:averagedNS_momentum}, which can be cumbersome numerically, and which simplifies the spectral analysis of energy fluxes. 
The resulting formulation reads
\begin{equation}
\label{eq:averagedNS_vorticity}
    \begin{split}
        \dot \omega = -\{\psi,\omega\} + \nu \Delta \omega + 2\nu \omega -\gamma \omega,\\
        -\Delta(1-\alpha^2 \Delta)^\beta \psi = \omega,
    \end{split}
\end{equation}
where $\omega \in C^\infty_0(\IS^2)$ is the vorticity of the fluid, and $\psi \in C^\infty(\IS^2)/\R$ is the stream function, which is determined up to a constant. 
The Poisson bracket $\{\cdot,\cdot\}$ is defined by
\begin{equation}
    \label{eq:poisson}
    \{f,g\} = \nabla f \cdot \nabla^\perp g,
\end{equation}
and its unaltered appearance in \eqref{eq:averagedNS_vorticity} alludes to the geometric properties of \al-modelling \cite{holm1998euler}.
The \ab-averaged Navier--Stokes equations \eqref{eq:averagedNS_vorticity} are a generalization of the usual Navier--Stokes equations, which are obtained as $\alpha = 0$. 
A qualitative illustration of the averaging is given in \cref{fig:vorticity_eyecandy}, depicting the smoothed vorticity fields for several values of \al~and \be.

There are several possible notions of energy and enstrophy in the averaged Navier--Stokes equations. 
However, a natural choice is to consider the energy and enstrophy that are preserved in the limit of vanishing viscosity and friction.
This energy is given by 
\begin{equation}
    \label{eq:energy}
    E = \frac{1}{2} \int_{\IS^2} \omega \psi 
\end{equation}
and the enstrophy is given by
\begin{equation}
    \label{eq:enstrophy}
    \cE = \frac{1}{2} \int_{\IS^2} \omega^2.
\end{equation}
In fact, the \ab-averaged Euler equations, given by 
\begin{equation}
    \label{eq:averagedEuler}
    \begin{split}
        \dot \omega = -\{\psi,\omega\},\\
        -\Delta(1-\alpha^2 \Delta)^\beta \psi = \omega,
    \end{split}
\end{equation}
form a Hamiltonian system on $C^\infty_0(\IS^2)$ with Hamiltonian $ H = \int_{\IS^2} \omega \psi$. The smoothed stream function enters the definition of the energy, thereby inhibiting the creation of scales below a certain threshold defined by \al~and yielding a nonlinear dispersive modification of the Navier--Stokes equations \cite{foias2001navier}. 

For the \al-model, the parameter $\alpha$ profoundly affects the distribution of energy across the resolvable scales of motion \cite{lunasin2008spectral}. 
The energy scalings in spectral space of the \ab-averaged Navier--Stokes equations are the focus of the current study. 
In three dimensions, the energy spectrum of the NS-\al~model shows a $-3$ scaling at wave numbers where the averaging is dominant, rather than the standard $-5/3$ scaling \cite{foias2001navier}. 
A similar result has been established for the two-dimensional NS-\al~model theoretically and numerically, yielding a $-5$ scaling in the direct enstrophy cascade regime at wave numbers where the averaging is strong \cite{lunasin2007study}. 
Furthermore, the latter showed that the cascading conserved quantity in the \al-model determines the scaling of derived statistical quantities in two-dimensional turbulence.

The suppression of small-scale motions in averaged turbulence is computationally appealing, since this property facilitates numerical simulations. This property has been exploited in computational large-eddy simulation (LES) studies. Re-writing the averaged equations of motion in terms of the smoothed momentum casts them into a standard LES template, where the divergence of the turbulent stress tensor is modeled as a forcing term \cite{geurts2002alpha}. The nonlinearly dispersive \al-model, as opposed to commonly used dissipative models, encompasses Leray regularization \cite{leray1934mouvement, geurts2006leray} and has shown favorable results for turbulent mixing when compared to eddy-viscosity models \cite{geurts2002alpha, geurts2003regularization}. The \al-modelling approach has also found meaningful applications in ocean modelling, where transport phenomena often play an important role. The \al-model and Leray model are studied in the context of the primitive equation ocean model \cite{hecht2008lans, hecht2008implementation}, yielding flow statistics resembling those obtained with no-model high-resolution simulations. In addition, qualitative flow dynamics also complied with higher-resolution numerical simulations. The present focus is the distribution of energy in two-dimensional homogeneous isotropic turbulence on the sphere, which is studied in the context of the \ab-Navier--Stokes equations. The novel inclusion of \be~describes the rate at which high-frequency flow components are curbed, and thereby enables stricter control over the roll-off of the energy spectrum. Furthermore, a range of numerical experiments are carried out to study the robustness of the \ab-model and investigate at which parameter regimes the large-scale flow features can still be predicted reliably.

The Lagrangian-averaged Navier--Stokes equations share mathematical and statistical properties with the standard Navier--Stokes equations. 
For example, the \al-Navier--Stokes equations can be cast in a variational formulation \cite{marsden2003anisotropic}, using the assumption that fluctuations smaller than length scale \al~are frozen in the mean flow. 
In three-dimensional turbulence, the \al-model is shown to not affect the connection between second- and third-order structure functions at separation distances larger than \al~\cite{holm2002karman}, further substantiating that the model is suited to accurately resolve the large scales of turbulent motion. 
Moreover, a derivation of the NS-\al~equations from the Kelvin circulation theorem is given by taking the integral over a loop moving with a spatially filtered velocity \cite{foias2001navier}. 
For two-dimensional fluids, this implies that inviscid averaged equations retain the conserved quantities of the two-dimensional Euler equations. 
The \ab-NS model may thus serve as a deterministic reference model for flow descriptions in which the transport velocity is stochastically perturbed, such as (Lagrangian-averaged) stochastic advection by Lie transport ((LA-)SALT) \cite{Arnaudon2012,holm2015variational, drivas2020lagrangian}.


In this paper, we make theoretical predictions about the transfer and cascade of energy and enstrophy in the \ab-NS equations on the sphere, and we verify these predictions numerically
using Zeitlin's structure-preserving spatial discretization of the Euler equations \cite{zeitlin1991finite, zeitlin2004self}. 
This computational method replaces the Hamiltonian system on the Poisson algebra of smooth zero-mean functions on the sphere with a Hamiltonian system on the Lie algebra of skew-Hermitian matrices of size $N \times N$. 
In particular, as the mathematical structure of the continuous system and the discrete system are geometrically and algebraically similar, we expect that the spectral behavior of the continuous system is accurately captured by the discrete system even at modest resolutions. This has been demonstrated for homogeneous isotropic turbulence on the sphere, where Kraichnan's double cascade \cite{kraichnan1967inertial} is accurately captured in Zeitlin-based numerical simulations \cite{cifani2022casimir}.

The paper is structured as follows. In \cref{sec:cascadedirs} we apply the analysis of Lindborg and Nordmark~\cite{lindborg2022two} to derive the cascade directions and spectrum scaling rates for averaged turbulence.
This is followed by numerical experiments in \cref{sec:experiments}. 
We verify the results of \cref{sec:cascadedirs} in two experiments, to study the averaging effect on both cascades. The paper is concluded in \cref{sec:conclusion}.  

\section{Cascade directions and spectrum scalings}
\label{sec:cascadedirs}
In this section, we provide arguments for the transfer and cascade of energy and enstrophy for the \ab-averaged Navier--Stokes equations \eqref{eq:averagedNS_vorticity}. 

We show that the double cascade of two-dimensional turbulence also holds for the averaged equations. That is, energy and enstrophy below a forcing wave number move from low to high wave numbers, and in the opposite direction above the forcing wave number, as predicted by Kraichnan \cite{kraichnan1967inertial}. 
Briefly, this is because the definitions of the energy, enstrophy and the corresponding rates of change in the NS-\al~model \eqref{eq:averagedNS_vorticity} equals the spherical two-dimensional case up to a dimensionless wave number-dependent factor, since the averaging operator $(1-\alpha^2 \Delta)^\beta$ is diagonalized by the spherical harmonics basis. 
The key is to identify in which expressions this scaling factor appears, and how it affects the energy and enstrophy transfer rates.

We derive these expressions in detail by expanding the relevant physical quantities in the spherical harmonic basis, following the approach of Lindborg and Nordmark \cite{lindborg2022two}. 
To this end, let $(Y_{l,m}, l \in \N_0, m=-l,\ldots,l)$ be the \emph{spherical harmonic functions}, i.e., the eigenfunctions of the Laplace--Beltrami operator~$\Delta$. 
The corresponding eigenvalues $-\lambda_{l}$ are given by
\begin{equation}
	 \Delta  Y_{l,m} = -l(l+1) Y_{l,m} \coloneq -\lambda_l Y_{l,m}.
\end{equation}

Consequently, a real scalar function $g\in L^2(\IS^2)$ can be expanded as  
\begin{align}
    g(\theta,\phi) = \sum_{l=0}^\infty g_l(\theta,\phi),
\end{align}
where $g_l(\theta,\phi) = \sum_{m = -l}^l a_{l,m}Y_{l,m}(\theta,\phi)$ 
with  $a_{l,m}= \langle g,Y_{l,m} \rangle_{L^2(\IS^2)}$. 
Since the stream function $\psi$ has the expansion 
\begin{align}
    \psi = \sum_{l = 0}^\infty \psi_l,
\end{align}
it follows from the relation $-\Delta(1-\alpha^2 \Delta)^\beta \psi = \omega$ that
\begin{align}
    \omega = \sum_{l = 0}^\infty \lambda_l (1+\alpha^2 \lambda_l)^\beta \psi_l \coloneqq  \sum_{l = 0}^\infty \omega_l,
    \label{eq:omega_expansion}
\end{align}
or, in other words, that $\psi_l = \lambda_l^{-1} (1+\alpha^2 \lambda_l)^{-\beta}\omega_l$. 
Consequently, the energy is expressed as 
\begin{align*}
    \mathsf E  = \frac{1}{2}\int_{\IS^2} \omega \psi = \sum_{l=0}^\infty \frac{\lambda_l^{-1} (1+\alpha^2 \lambda_l)^{-\beta} }{2}\langle \omega_l, \omega_l \rangle_{L^2(\IS^2)}. 
\end{align*}
We set $\frac{\lambda_l^{-1}(1+\alpha^2 \lambda_l)^{-\beta}}{2}\langle \omega_l, \omega_l \rangle_{L^2(\IS^2)}  =  \mathsf E(l)$ to be the energy in mode $l$. 
Furthermore, $\mathsf E$ is a natural definition of the energy since it
\begin{enumerate}
    \item is the Hamiltonian of the \ab~Euler equations;
    \item is a conserved quantity in the case of vanishing viscosity and friction. 
\end{enumerate}
In the limit of vanishing viscosity and friction, there are an infinite number of conserved quantities called \emph{Casimirs}, given by integrals of analytic functions of the vorticity.
All Casimirs can thus be expanded in Casimir functions of the form $\int_{\IS^2} \omega^k$, where $k \in \N$. 
A well-known Casimir is the \emph{enstrophy}, given by 
\begin{align*}
    \cE = \frac{1}{2}\int_{\IS^2} \omega^2 = \sum_{l= 0 }^\infty \frac{1}{2} \langle \omega_l,\omega_l \rangle_{L^2(\IS^2)}.
\end{align*}
Here, we use the series expansion \eqref{eq:omega_expansion} to obtain the right-hand side. The enstrophy in mode~$l$ is given by 
\begin{align*}
    \cE(l) = \lambda_l(1+\alpha^2 \lambda_l)^\beta \mathsf{E}(l). 
\end{align*}

With these definitions at hand, we turn our attention to the corresponding rates of change. 
\Cref{eq:averagedNS_vorticity} induces an evolution of $\mathsf{E}(l)$, which is found by multiplying \eqref{eq:averagedNS_vorticity} by $\omega_l$ and integrating over $\IS^2$,
\begin{align}
    \int_{\IS^2} \dot \omega \omega_l = -\int_{\IS^2} \{\psi,\omega\} \omega_l  +\int_{\IS^2} \left(\nu(\Delta \omega + 2\omega) - \gamma \omega \right) \omega_l \label{eq:NS_E_ev_1}. 
\end{align}
Note that we also have
\begin{align*}
    \int_{\IS^2} \dot \omega \omega_l = \frac{1}{2} \frac{\mathrm d}{\mathrm d t }\langle \omega_l,\omega_l \rangle = \dot{\cE}(l) = \lambda_l(1+\alpha^2 \lambda_l)^\beta \dot{\mathsf E}(l).
\end{align*}
We follow Lindborg and Nordmark \cite{lindborg2022two} and introduce the quantity 
\begin{align} \label{eq:Tl}
    T(l) = -\int_{\IS^2} \{\psi,\omega\} \psi_l,
\end{align}
so that the nonlinear term in \eqref{eq:NS_E_ev_1} is given by 
\begin{align}
   -\int_{\IS^2} \omega_l \{\psi,\omega\} =  \lambda_l(1+\alpha^2 \lambda_l)^\beta T(l).
\end{align}
\Cref{eq:Tl} reveals that the behavior of $T(l)$ is intimately connected to the Lie algebra structure of the Poisson bracket \eqref{eq:poisson}, and 
that the \emph{triadic interactions}, central in the analysis of Kraichnan \cite{kraichnan1967inertial}, Fj\o rtoft \cite{fjortoft1953changes} and Lindborg and Nordmark \cite{lindborg2022two}m 
are captured by the \emph{structure constants} of the Poisson algebra in the spherical harmonics basis. 
The geometric analysis of the triadic interactions warrants its own study and is not further investigated here.

The remaining terms in \eqref{eq:NS_E_ev_1} are rewritten as 
\begin{align}
    \int_{\IS^2} \omega_l \Delta \omega =& -\sum_{n=1}^\infty \lambda_n  \int_{\IS^2} \omega_l  \omega_n =   -2\lambda_l^2 (1+\alpha^2 \lambda_l)^\beta \mathsf E(l), \\
    \int_{\IS^2} \omega_l  \omega =& \sum_{n=1}^\infty   \int_{\IS^2} \omega_l  \omega_n =2\lambda_l(1+\alpha^2 \lambda_l)^\beta\mathsf E(l)
\end{align}
where we have used that $\Delta \omega_l = -\lambda_l \omega_l$.
Thus, the energy transfer equation  for the \ab-averaged Navier--Stokes equations reads
\begin{align}
    \dot{\mathsf E}(l) = T(l) - 2\nu(\lambda_l \mathsf E(l) - 2\mathsf E(l)) -\gamma \mathsf E(l).
    \label{eq:NS_e_ev}
\end{align}
Note that this is the same equation as for the (standard) Navier--Stokes equations. 
To continue the analysis, we need to study $T(l)$. 
Observe that $\psi$ and $\omega$ may be expanded into their respective modes, so that $T(l)$ is rewritten as
\begin{align}
    T(l) = -\sum_{n=0}^\infty \sum_{s= 0 }^\infty \int_{\IS^2} \psi_l \{\psi_n,\omega_s\}.
    \label{eq:T_expansion}
\end{align}
By noting that $\psi_l$, $\psi_n$ and $\omega_s$ all consist of finite linear combinations of spherical harmonics, it becomes evident from \eqref{eq:T_expansion} that the qualitative behavior of $T(l)$ is described by the structure constants $C_{lm,nm',sm''} = \int_{\IS^2} Y_{l,m} \{Y_{n,m'},Y_{s,m''}\}$. In particular, this is independent of the values of \al~and \be. Consequently, the analysis of the energy transfer in the \ab-averaged Navier--Stokes equations is identical to the analysis of the usual Navier--Stokes equations on the sphere, detailed by Lindborg and Nordmark \cite{lindborg2022two}, up to the scaling factor $(1+\alpha^2 \lambda_l)^\beta$.

We now show that the kinetic energy dissipation vanishes for the averaged Navier--Stokes equations in the limit of small viscosity, following the same analysis as Lindborg and Nordmark \cite{lindborg2022two}. This result implies that there can be no Richardson energy cascade, and that energy is instead transferred to large scales in an inverse cascade.
The evolution of the average total energy $\bar E$ and enstrophy $\bar\cE$ are given by
\begin{align}
    &\frac{\dd \bar E}{\dd t}  = -\varepsilon \\
     &\frac{\dd \bar \cE}{\dd t} =  -\eta,
\end{align}
where $\varepsilon = \sum_{l=1}^\infty (2\nu\lambda_l-4\nu+2\gamma)\mathsf E(l)$ and $\eta = \sum_{l=1}^\infty \lambda_l (1+\alpha^2 \lambda_l)^\beta(2\nu\lambda_l-4\nu+2\gamma)\mathsf E(l)$ respectively denote the average energy and enstrophy dissipation rates.

By omitting energy and enstrophy loss due to friction, i.e., terms with $\gamma$, the result of Lindborg and Nordmark \cite{lindborg2022two} can be cast in the present notation as \begin{align}
    &\varepsilon = \sum_{l=1}^\infty 2\nu\left(\lambda_l - 2\right)\mathsf E(l), \\
    &\eta = \sum_{l=1}^\infty 2\lambda_l(1+\alpha^2 \lambda_l)^\beta\nu\left(\lambda_l - 2\right) \mathsf E(l).
\end{align}
We deduce that kinetic energy dissipation vanishes in the limit $\nu\to 0$.
Indeed, inserting $\lambda_l^{-1} (1+\alpha^2 \lambda_l)^{-\beta} \cE(l) = \mathsf E(l)$ into the equation of $\varepsilon$ yields \begin{equation}
    \varepsilon = \sum_{l=1}^\infty 2\nu\left(\lambda_l - 2\right) \lambda_l^{-1}(1+\alpha^2 \lambda_l)^{-\beta} \cE(l). 
\end{equation}
Since  \begin{equation}
    \frac{\lambda_l - 2}{\lambda_l(1+\alpha^2 \lambda_l)^\beta} < 1,
\end{equation}
for all $l$, we have that 
\begin{equation}
    \varepsilon = \sum_{l=1}^\infty 2\lambda_l^{-1}  (1+\alpha^2 \lambda_l)^{-\beta} \nu\left(\lambda_l - 2\right)  \cE(l) < \sum_{l=1}^\infty 2\nu  \cE(l) = 2\nu  \bar \cE,
\end{equation}
The enstrophy is a positive quantity that is non-increasing in time, since $\eta$ is nonnegative. This indicates that the kinetic energy dissipation has an upper bound in terms of the viscosity and the total enstrophy. The enstrophy is bounded from above; hence the energy dissipation must vanish if $\nu\to 0$. (It is clear that the kinetic energy dissipation cannot be negative.) This result does not follow in three-dimensional turbulence, where the enstrophy can grow due to vortex stretching \cite{boffetta2012two}.

We continue the analysis by studying the spectral energy and enstrophy fluxes. These are respectively defined as 
\begin{align}
    &\Pi_{\mathsf E}(k) = \sum_{l = k}^\infty T(l)  = -\sum_{l = 1}^{k-1} T(l), \label{eq:energy_flux}\\ 
    &\Pi_\mathcal{E}(k) = \sum_{l = k}^\infty \lambda_l (1+\alpha^2 \lambda_l)^\beta T(l)  = -\sum_{l = 1}^{k-1} \lambda_l (1+\alpha^2 \lambda_l)^\beta T(l), \label{eq:enstrophy_flux}
\end{align}
where energy and enstrophy conservation is used to obtain the right-hand sides \cite{lindborg2022two}. 
Now, the analysis of the fluxes presented by Lindborg and Nordmark \cite{lindborg2022two} can be repeated verbatim since the behavior of the fluxes is determined by the structure constants, which are independent of \al~and \be.
The only difference is that the scaling factor describing the enstrophy transfer within a triad of modes, when given in terms of the energy transfer within these modes, is given by $\lambda_l (1+\alpha^2 \lambda_l)^\beta$ instead of $\lambda_l$.
Thus, the qualitative reasoning in Lindborg and Nordmark \cite[Section 5-6]{lindborg2022two} can still be applied. We therefore have that the enstrophy flux is positive when the energy flux is zero, and vice versa for the energy flux.
Moreover, in the case of narrow-band forcing in spectral space, more energy is transferred to low wave numbers than to high wave numbers, and the opposite holds true for the enstrophy \cite{gkioulekas2007new}. 

Assume now that the forcing is concentrated at a wave number $l_f$, such that $ l_f \ll 1/\alpha$. 
In that case, we expect three distinct scaling regimes to appear in the energy spectrum.
At wave numbers $l \ll 1/\alpha $, i.e., wave numbers at which the effect of the averaging is negligible, the distribution of energy should follow the double cascade observed in the standard Navier--Stokes equations. That is, the energy spectrum scales as $l^{-5/3}$ in the energy cascade range (when $l \ll l_f $) and as $l^{-3}$ in the enstrophy cascade range (when $l_f \ll l \ll 1/\alpha$).
This means that the large flow scales are largely unaffected by the averaging.
A third regime appears at wave numbers such that $l \gg 1/\alpha$, i.e., wave numbers at which the effect of the averaging is significant.
In this regime, dimensional analysis as in Kraichnan \cite{kraichnan1967inertial} and the approximation $l(l+1) \approx l^2$ for large wave numbers, as in Lindborg and Nordmark \cite{lindborg2022two}, predicts that 
\begin{align}
    \mathsf (1+\alpha^2 l^2)^\beta E(l) \approx l^{-3}.
\end{align}
Equivalently, this implies that the energy spectrum scales as 
\begin{align}
    E(l) \approx l^{-3-2\beta},
\end{align}
resulting in a steeper decline of the energy than predicted in the enstrophy cascade range, with a stronger suppression of energy as \be~increases.
This is in line with Lunasin et al. \cite{lunasin2007study}, who derive via dimensional analysis and numerical simulations that the energy spectrum should scale as $l^{-5}$ for $\beta = 1$. 
\begin{remark}
Note that the above derivation is not limited to the operator $(1-\alpha^2\Delta)^\beta$. Indeed, let $g = \sum_{l = 0}^\infty g_l \in C^\infty(\IS^2)$. 
If $L:C^\infty(\IS^2) \mapsto C^\infty(\IS^2)$ is a linear operator defined by 
\begin{align*}
    L(g) = \sum_{l=0}^\infty \tilde{L}(\lambda_l) g_l
\end{align*}
for some function $\tilde{L}$ such that 
\begin{enumerate}
    \item  $\tilde{L}(\lambda_l)$ is a dimensionless quantity, 
    \item $\frac{\lambda_l}{ \tilde{L}(\lambda_l)} < 1$,
\end{enumerate}
then the above analysis also holds when $\psi = \tilde{L}(\omega)$. 
For instance, $\tilde{L}(\lambda_l)$ could be chosen such that it equals $\lambda_l$ if $l$ is small, and only apply \ab~smoothing for wave numbers above a specified value.  
\end{remark}

\section{Numerical experiments} \label{sec:experiments}
We now move on to numerical demonstrations of the results derived in \cref{sec:cascadedirs}.

\subsection{Computational method}
The spatial discretization of the equations of motion \eqref{eq:averagedNS_vorticity} follows the self-consistent finite-dimensional truncation method of Zeitlin \cite{zeitlin2004self}.
In the absence of viscous dissipation and external forcing, the vorticity equation \eqref{eq:averagedNS_vorticity} forms a Lie--Poisson system \cite{arnold2009topological, marsden2013introduction} on the space of smooth functions on the sphere $\mathcal{C}^\infty(S^2)$.
The Zeitlin approach provides a finite-dimensional truncation of the Poisson bracket through geometric quantization \cite{hoppe1989diffeomorphism, bordemann1991gl, bordemann1994toeplitz}, thereby enabling a fully structure-preserving discretization of the convection operator.
The method relies on a projection $\Pi_N:\mathcal{C}^\infty(S^2)\to \mathfrak{u}(N)$ of smooth functions to complex $N\times N$ skew-Hermitian matrices and replacing the Poisson bracket by the matrix commutator $\frac{1}{\hbar}[\cdot, \cdot]$ for $\hbar = 2/\sqrt{N^2-1}$.
A spherical harmonic basis element $Y_{lm}$ is then identified with a matrix basis $T_{lm}^N\in \mathfrak{u}(N)$.
The value $N$ can be thought of as the adopted numerical resolution, and we have for $N\to\infty$:
\begin{enumerate}
    \item $\Pi_N f - \Pi_N g \to 0 \implies f=g$,
    \item $\lVert \Pi_N \{f, g\} - \frac{1}{\hbar} \left[ \Pi_N f, \Pi_N g\right]\rVert_\infty = \mathcal{O}(1/N)$,
    \item $\Pi_N Y_{lm} = T_{lm}^N$,
\end{enumerate}
where $\lVert\cdot\rVert_\infty$ denotes the spectral norm.
The Laplacian operator $\Delta$ is replaced by the discrete Laplacian operator $\Delta_N$ of Hoppe and Yau~\cite{hoppe1998some}. 
The critical property of $\Delta_N$ is that the basis elements $T_{lm}^N$ are the eigenvectors with corresponding eigenvalues $-l(l+1)$, analogous to the Laplacian and the spherical harmonics. 

A discrete approximation of the vorticity $\omega$ is obtained after identifying a spherical harmonic function $Y_{lm}$ with the matrix harmonic $T_{lm}^N$. 
The spherical harmonic coefficients of the vorticity are computed as $c_{lm}(t) = \langle \omega(t), Y_{lm}\rangle_{L^2(S^2)}$, after which the matrix approximation $W$ of $\omega$ reads
\begin{equation}
    W(t) = \sum_{l=0}^{N-1}\sum_{m=-l}^l c_{lm}(t) T_{lm}^N.
\end{equation}
The stream matrix $P$ is computed analogously from $\psi$. 
The discrete approximation to the equations of motion \eqref{eq:averagedNS_vorticity} are given by \begin{equation} \label{eq:discrete_averagedNS_vorticity}
\begin{split}
    \dot{W} = -\frac{1}{\hbar}[P, W] + \nu\Delta_N W + 2\nu W -\gamma W, \\
    -\Delta_N(1-\alpha^2\Delta_N)^\beta P = W.
\end{split}
\end{equation} 
In particular, the system \eqref{eq:discrete_averagedNS_vorticity} has Casimir functions in the absence of viscosity and external forcing and damping. 
These are defined as the traces of the powers of vorticity \cite{zeitlin1991finite}, \begin{equation}
    C_k(W) = \trace (W^k),
\end{equation}
analogous to the integrated powers of vorticity in the original system.
We employ an isospectral Lie--Poisson integrator \cite{modin2020lie, modin2020casimir, cifani2023efficient} for the convective term to preserve these Casimir functions. 
The viscous dissipation and external forcing and damping terms are integrated using a Crank--Nicolson scheme \cite{crank1947practical}. 
We denote these time integration schemes over a time interval of length $h$ by $\phi_{\mathrm{iso}, h}$ and $\phi_{\mathrm{CN}, h}$, respectively. 
A single time integration step of length $h$ for \eqref{eq:discrete_averagedNS_vorticity} is then carried out using the Strang splitting \cite{strang1968construction} as \begin{equation}
    W^{n+1} = \left( \phi_{\mathrm{CN}, h/2} \circ \phi_{\mathrm{iso}, h} \circ \phi_{\mathrm{CN}, h/2} \right)(W^n),
\end{equation}
where the superscript $n$ denotes the $n^\mathrm{th}$ time step. 
This provides a second-order time discretization of the dynamics.

\subsection{Details of the numerical experiments}
Two series of numerical experiments are carried out at resolution $N=512$. 
All results are compared to numerical results of the standard Navier--Stokes equations obtained with the same computational method. 
Each numerical realization is driven to a statistically stationary state via an external forcing placed at a forcing wave number $l_f$, combined with a friction coefficient $\gamma$ to prevent energy build-up in low wave number modes. 
A single Reynolds number is chosen for both sets of simulations, where we adopt the forcing-dependent definition $\mathit{Re}=\left[E(l_f)/l_f\right]^{1/2}/\nu$ as presented by Kraichnan \cite{kraichnan1967inertial}. 
The external forcing is defined as white noise in time to control the energy injection rate \cite{boffetta2007energy}, multiplied by a magnitude $f_\mathrm{mag}$. 
We estimate $E(l_f)$ as $E(l_f)=f_\mathrm{mag}^2 (2 l_f + 1) / (\lambda_{l_f}(1+\alpha^2\lambda_{l_f})^\beta)$ and adjust the viscosity parameter $\nu$ so that a specified Reynolds number is obtained. 
All numerical simulations are run until a statistically stationary state is reached, which we define by having a constant time-averaged energy spectrum over different time intervals.

In the first set of experiments, the forcing is placed at wave number $l_f=5$ to allow the maximal possible development of the enstrophy inertial subrange. The Reynolds number is set to 1200. 
A sequence of values for \al~and \be~is applied in separate numerical simulations to obtain clear numerical evidence of the different scaling regimes predicted in the previous section. 
Here, we adopt all combinations of $\alpha\in\left\{1/100, 1/50, 1/20 \right\}$ and $\beta\in\left\{1, 2\right\}$. 

The second set consists of simulations in which a double cascade can develop, by placing the external forcing at wave number $l_f=50$ and setting the Reynolds number at 600. 
We demonstrate that this choice of forcing already allows for the formation of a double cascade at the employed resolution. 
The same values of \al~and \be~are applied as in the first set of experiments to study the effects of the averaging procedure in both regimes of the classical double cascade. 

\subsection{Enstrophy cascade scalings}
\begin{figure}[h!]
    \centering
    \includegraphics[width=0.48\linewidth]{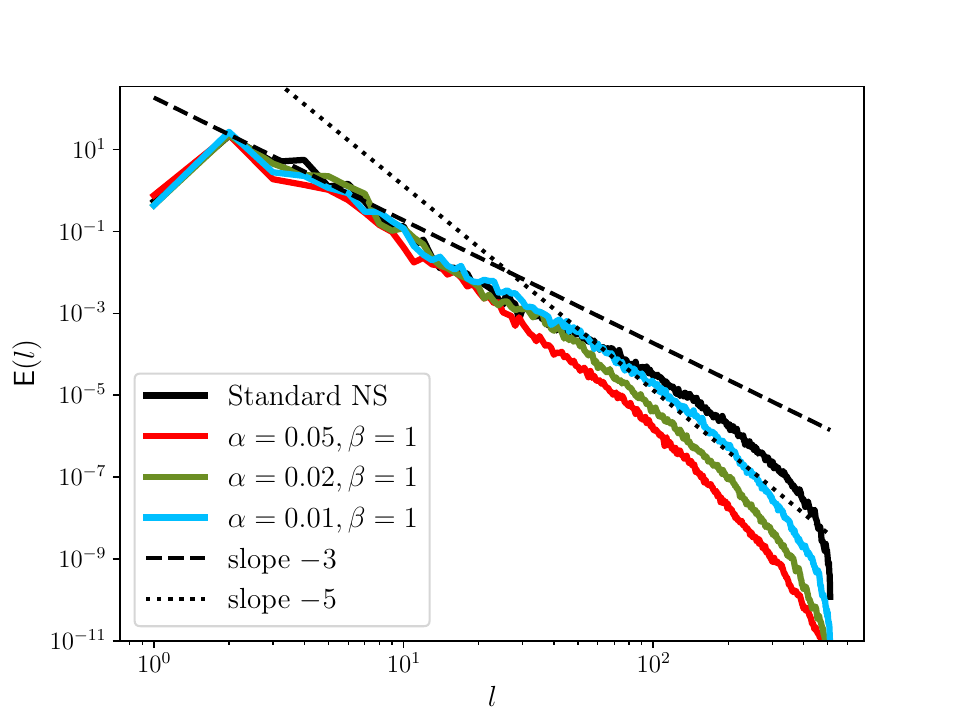}
    \includegraphics[width=0.48\linewidth]{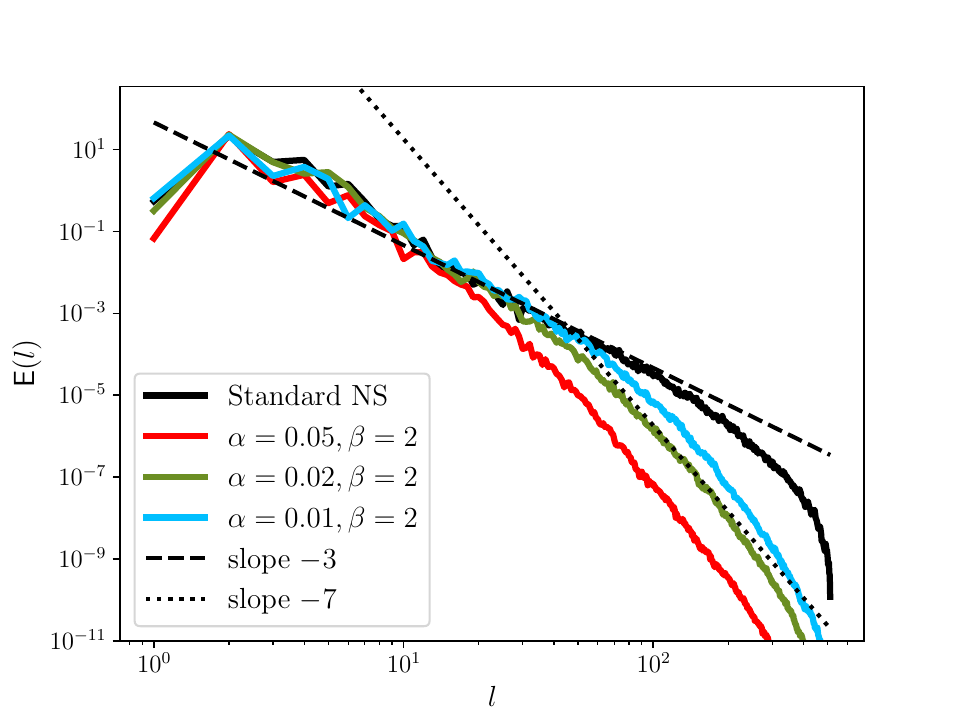}
    \caption{Energy spectra as a function of the wave number for various choices of \al~and \be. The Kraichnan scaling $-3$ and the expected scalings $-3-2\beta$ are provided for reference.}
    \label{fig:spectra_enstrophy_cascade}
\end{figure}

The first set of numerical simulations is used to verify the predicted scaling laws in the enstrophy cascade regime for different values of \al~and \be. 
The energy spectra are shown in \cref{fig:spectra_enstrophy_cascade} along with reference scalings. 
The standard Navier--Stokes result is the same in both panels, and the departures from this energy spectrum clearly display the effect of the averaging on the distribution of energy across the scales of motion.

The left panel shows the results for $\beta=1$. 
The predicted $-5$ scaling is attained in the results of the averaged Navier--Stokes equations. 
This is best observed for $\alpha=1/20$, which causes the averaging term to become dominant at wave numbers well separated from wave numbers at which dissipative effects are strong. 
The averaging is further amplified for $\beta=2$, visible from the energy spectra in the right panel. 
A clear $-7$ scaling is found for these parameter values, which agrees with the prediction in the previous section. 
The cusp between the two scalings in the energy spectrum appears sharper than for $\beta=1$, which is due to the term $(1-\alpha^2\lambda_l)^\beta$ growing exponentially for increasing \be~and thus strongly increasing the averaging effect.

\subsection{Robustness of the double cascade and spectral fluxes}

\begin{figure}[h!]
    \centering
    \includegraphics[width=0.48\linewidth]{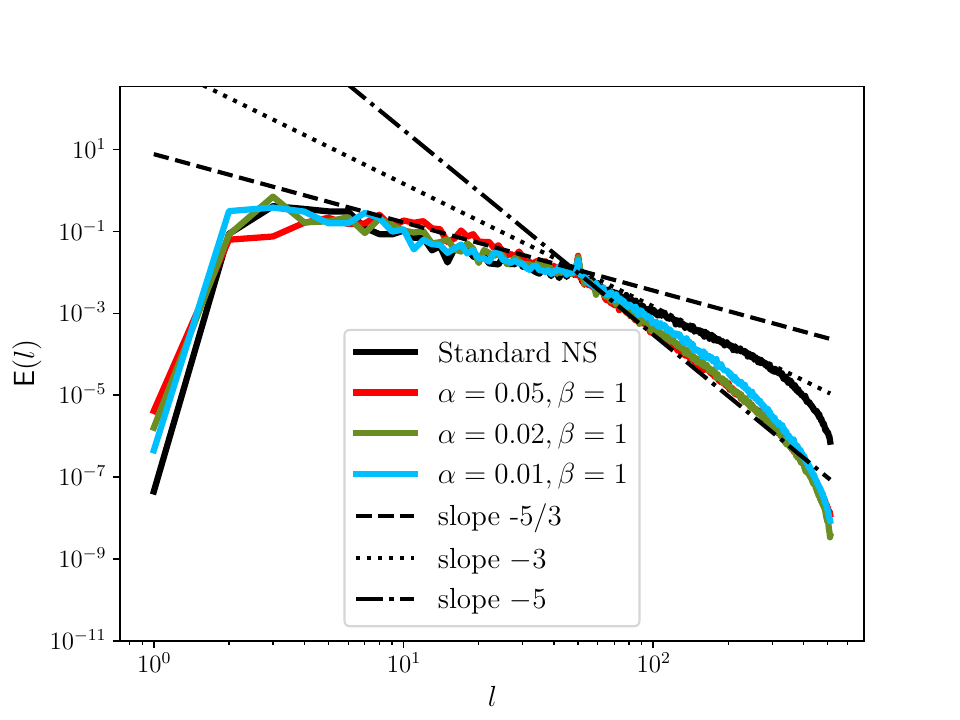}
    \includegraphics[width=0.48\linewidth]{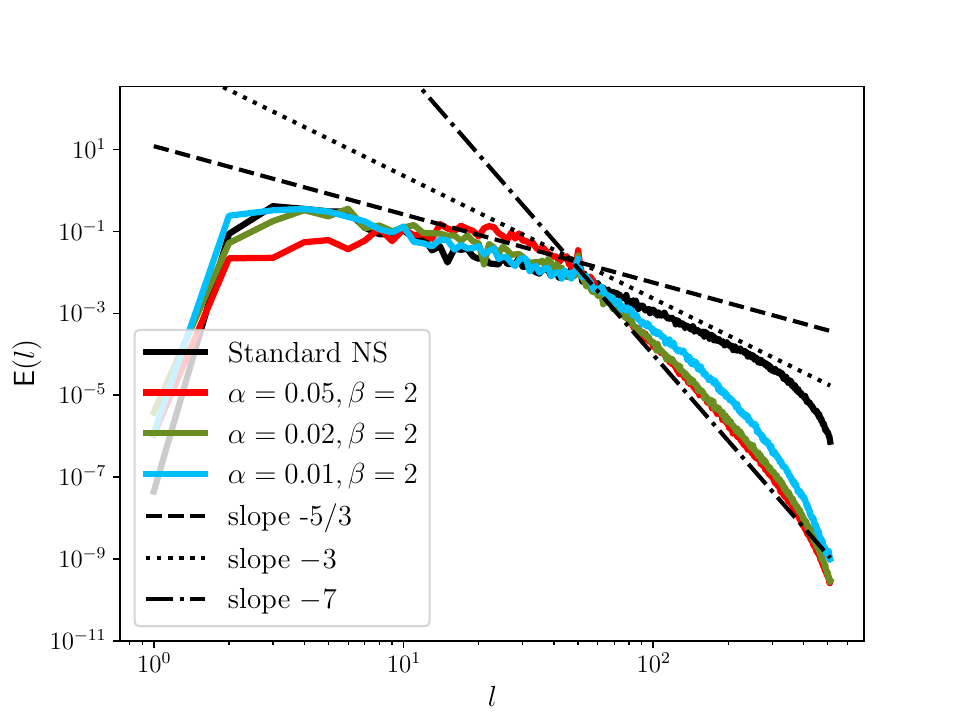}
    \caption{Energy spectra as a function of the wave number for various choices of \al~and \be. The Kraichnan scalings $-5/3$ and $-3$. The expected scalings $-3-2\beta$ are provided for reference.}
    \label{fig:spectra_double_cascade}
\end{figure}

In the second set of numerical simulations, we investigate how the averaging affects the double cascade observed in two-dimensional turbulence. 
The corresponding energy spectra are given in \cref{fig:spectra_double_cascade}, where the spectrum of the standard Navier--Stokes equations is the same in both panels for comparison.
Even at the modest adopted resolution $N=512$, the $-5/3$ and $-3$ scalings as predicted by Kraichnan are accurately captured. 
The \ab-averaged Navier--Stokes equations display the expected scalings at high wave numbers for all considered values of \al~and \be. 
This coincides with the results reported in the previous subsection. 
Of particular interest are the results obtained with $\alpha = 1/20$ ($\alpha=0.05$), where the term $\alpha^2 \lambda_l$ becomes significant at wave numbers smaller than the forcing wave number $l_f = 50$. 
For this value of \al, the averaging effect for $\beta=1$ is difficult to discern at wave numbers $l<50$. 
However, a deviation is observed in the inverse energy cascade for $\alpha=1/20$ and $\beta=2$. 

Injecting the energy at wave number $l_f=50$ allows for a clear identification of the energy and enstrophy fluxes. 
The energy is expected to move from the forcing wave number towards low wave numbers in the standard Navier--Stokes equations \cite{kraichnan1967inertial, lindborg2022two}, whereas the enstrophy is expected to move to high wave numbers. 
This qualitative behavior is not expected to change for the \ab-Navier--Stokes equations, following the analysis in \cref{sec:cascadedirs}. 
To demonstrate this, we take a vorticity field from the standard Navier--Stokes equations in a statistically stationary state and compute the corresponding stream function for various values of \al~and \be. 
Subsequently, the spectral fluxes are computed following their respective definitions in \eqref{eq:energy_flux} and \eqref{eq:enstrophy_flux}. 

\begin{figure}
    \centering
    \includegraphics[width=0.96\linewidth]{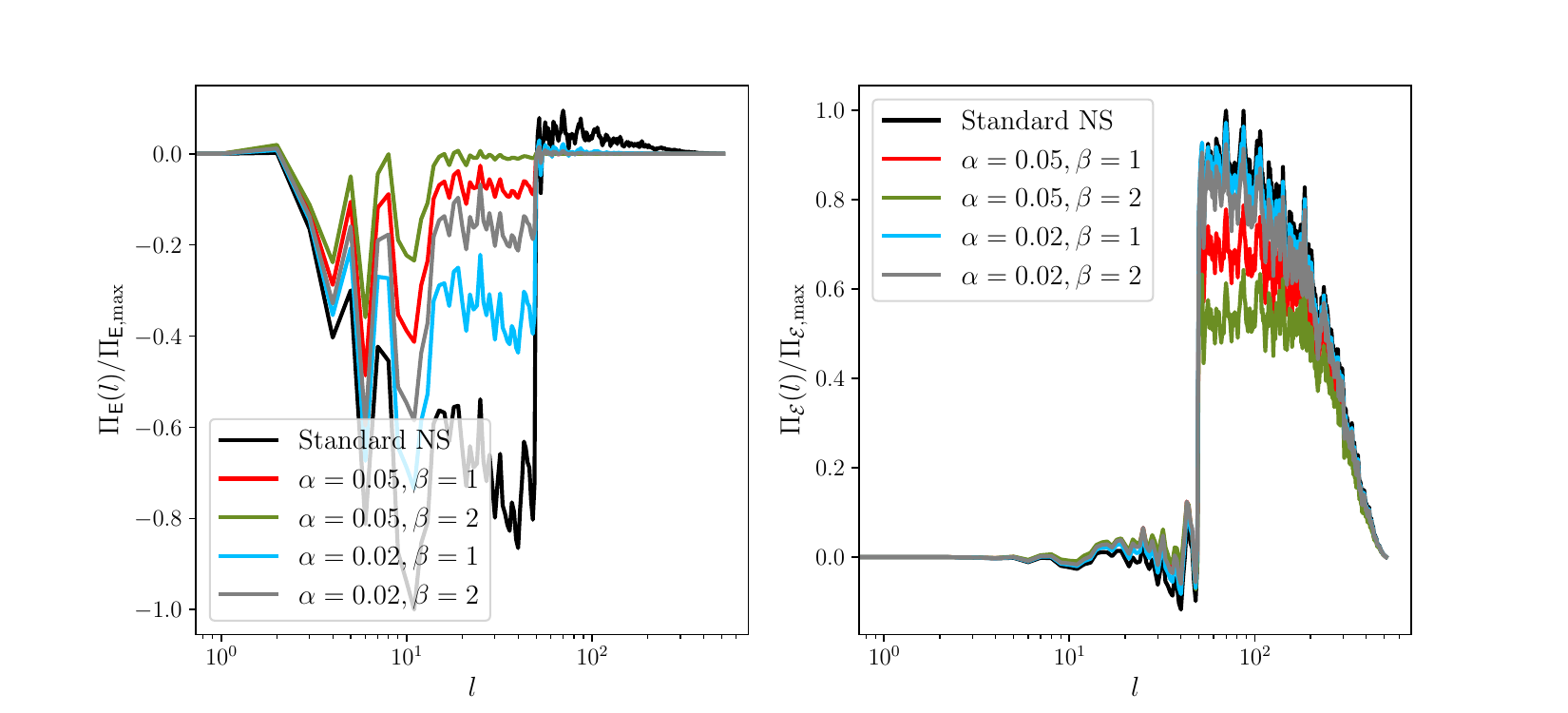}
    \caption{Spectral convective energy fluxes (left) and enstrophy fluxes (right) as a function of the wave number.}
    \label{fig:transfers}
\end{figure}

The fluxes are presented in \cref{fig:transfers}, normalized by the maximum absolute value of the corresponding flux for the standard Navier--Stokes equations. 
The forcing causes the jump observed in both fluxes at $l=50$. The negative energy flux for $l<l_f$ indicates energy moving to low wave numbers. 
Similarly, the positive enstrophy flux for $l>l_f$ implies that enstrophy is transferred to high wave numbers. 
This qualitative behavior is also observed in the fluxes of the \ab-Navier--Stokes equations, suggesting that the double cascade is still present in the averaged equations. 
After all, the transfer function $T(l)$ does not change qualitatively since it is defined in terms of the Poisson bracket and the vorticity, which both do not change when averaging, and the stream function, which is suppressed at higher wave numbers. 
This smoothing effect is especially visible as the energy fluxes approach zero at wave numbers $20<l<50$ for $\alpha=0.05$ and $\beta=2$. 
A less pronounced damping is visible in the enstrophy flux. 
The disparity between the damping of the energy flux and the enstrophy flux is readily explained by the appearance of the stream function in their respective definitions. 
Namely, the stream function is subject to the averaging operator, which dampens the component at wave number $l$ by a factor $(1+\alpha^2\lambda_l)^{-\beta}$. 
This factor appears twice in the definition of the energy flux and only once in the enstrophy flux. 
Thus, the smoothing has a larger effect on the former than on the latter. 
The different effects on each of these fluxes implies that one can control the relation of the fluxes through an appropriate choice of averaging operator.  

\section{Conclusions} \label{sec:conclusion}
In this paper, we carried out spectral analysis for averaged turbulence on the sphere.
The scaling laws for a class of Lagrangian-averaged Navier--Stokes equations were derived by extending the results of Lindborg and Nordmark \cite{lindborg2022two} and utilizing the geometry of two-dimensional fluids in two distinct ways.
First, the scaling laws were determined by studying the governing equations in stream function-vorticity formulation that yielded simple expressions for the relevant terms.
Second, the averaging operator was applied in a manner that retains the underlying geometric structure of the equations, thereby enabling an extension of previous results to encompass Lagrangian-averaged turbulence.

The averaging operator was found to suppress the energy in high wave number flow components, which is in agreement with previous studies on the Navier--Stokes-\al~model.
Here, an extension to previous models was included through an additional parameter that determines the number of times that the smoothing is applied.
This lead to a class of averaging operators that not only determines the onset of the energy suppression, but also allows for control over the rate of suppression.
In particular, a necessary condition for the existence of the inverse cascade in two-dimensional turbulence was derived for the averaging operator.
Furthermore, the energy flux and enstrophy flux were found to have a distinct dependence on the parameters of the averaging operator, allowing for control over these fluxes through the choice of parameters. 
The derived scaling laws, existence of the inverse cascade, and the energy and enstrophy fluxes were demonstrated in numerical experiments performed with a structure-preserving integrator for flows on the sphere.

The results presented in this paper can serve as a reference for spectral analysis of other global geophysical fluid models that possess similar geometric properties as the Euler equations, such as the quasi-geostrophic equations \cite{luesink2024geometric}.
Additionally, the current results can be compared to geophysical fluid models where the advection velocity is stochastically perturbed and a smoothing effect can be achieved by taking the expectation \cite{Arnaudon2012, holm2015variational, drivas2020lagrangian}.

\bibliographystyle{abbrv}
\bibliography{refs_abb}

\begin{thebibliography}{10}

\bibitem{Arnaudon2012}
M.~Arnaudon and A.~B. Cruzeiro.
\newblock Lagrangian {N}avier–{S}tokes diffusions on manifolds: Variational principle and stability.
\newblock {\em Bull. Sci. Math.}, 136(8):857–881, Dec. 2012.

\bibitem{arnold2009topological}
V.~I. Arnold and B.~A. Khesin.
\newblock {\em Topological methods in hydrodynamics}, volume~19.
\newblock Springer, 2009.

\bibitem{boffetta2007energy}
G.~Boffetta.
\newblock Energy and enstrophy fluxes in the double cascade of two-dimensional turbulence.
\newblock {\em J. Fluid Mech.}, 589:253--260, 2007.

\bibitem{boffetta2012two}
G.~Boffetta and R.~E. Ecke.
\newblock Two-dimensional turbulence.
\newblock {\em Annu. Rev. Fluid Mech.}, 44(1):427--451, 2012.

\bibitem{bordemann1991gl}
M.~Bordemann, J.~Hoppe, P.~Schaller, and M.~Schlichenmaier.
\newblock {gl}($\infty$) and geometric quantization.
\newblock {\em Comm. Math. Phys.}, 138:209--244, 1991.

\bibitem{bordemann1994toeplitz}
M.~Bordemann, E.~Meinrenken, and M.~Schlichenmaier.
\newblock Toeplitz quantization of {K}{\"a}hler manifolds ang gl (n), {N}→$\infty$ limits.
\newblock {\em Comm. Math. Phys.}, 165:281--296, 1994.

\bibitem{cifani2023efficient}
P.~Cifani, K.~Modin, and M.~Viviani.
\newblock An efficient geometric method for incompressible hydrodynamics on the sphere.
\newblock {\em J. Comput. Phys.}, 473:111772, 2023.

\bibitem{cifani2022casimir}
P.~Cifani, M.~Viviani, E.~Luesink, K.~Modin, and B.~J. Geurts.
\newblock Casimir preserving spectrum of two-dimensional turbulence.
\newblock {\em Phys. Rev. Fluids}, 7(8):L082601, 2022.

\bibitem{crank1947practical}
J.~Crank and P.~Nicolson.
\newblock A practical method for numerical evaluation of solutions of partial differential equations of the heat-conduction type.
\newblock In {\em Mathematical proceedings of the Cambridge philosophical society}, volume~43, pages 50--67. Cambridge University Press, 1947.

\bibitem{drivas2020lagrangian}
T.~D. Drivas, D.~D. Holm, and J.-M. Leahy.
\newblock Lagrangian averaged stochastic advection by lie transport for fluids.
\newblock {\em J. Stat. Phys.}, 179(5):1304--1342, 2020.

\bibitem{fjortoft1953changes}
R.~Fj{\o}rtoft.
\newblock On the changes in the spectral distribution of kinetic energy for twodimensional, nondivergent flow.
\newblock {\em Tellus}, 5(3):225--230, 1953.

\bibitem{foias2001navier}
C.~Foias, D.~D. Holm, and E.~S. Titi.
\newblock The {N}avier--{S}tokes-alpha model of fluid turbulence.
\newblock {\em Phys. D}, 152:505--519, 2001.

\bibitem{geurts2002alpha}
B.~J. Geurts and D.~D. Holm.
\newblock {\em Alpha-modeling strategy for {LES} of turbulent mixing}.
\newblock Springer, 2002.

\bibitem{geurts2003regularization}
B.~J. Geurts and D.~D. Holm.
\newblock Regularization modeling for large-eddy simulation.
\newblock {\em Phys. Fluids}, 15(1):L13--L16, 2003.

\bibitem{geurts2006leray}
B.~J. Geurts and D.~D. Holm.
\newblock Leray and lans-$\alpha$ modelling of turbulent mixing.
\newblock {\em J. Turbul.}, (7):N10, 2006.

\bibitem{gkioulekas2007new}
E.~Gkioulekas and K.~K. Tung.
\newblock A new proof on net upscale energy cascade in two-dimensional and quasi-geostrophic turbulence.
\newblock {\em J. Fluid Mech.}, 576:173–189, 2007.

\bibitem{hecht2008lans}
M.~Hecht, D.~Holm, M.~Petersen, and B.~Wingate.
\newblock The lans-$\alpha$ and leray turbulence parameterizations in primitive equation ocean modeling.
\newblock {\em J. Phys. A}, 41(34):344009, 2008.

\bibitem{hecht2008implementation}
M.~W. Hecht, D.~D. Holm, M.~R. Petersen, and B.~A. Wingate.
\newblock Implementation of the lans-$\alpha$ turbulence model in a primitive equation ocean model.
\newblock {\em J. Comput. Phys.}, 227(11):5691--5716, 2008.

\bibitem{holm2002karman}
D.~D. Holm.
\newblock {K}arman--{H}owarth theorem for the lagrangian-averaged {N}avier--{S}tokes--alpha model of turbulence.
\newblock {\em J. Fluid Mech.}, 467:205--214, 2002.

\bibitem{holm2015variational}
D.~D. Holm.
\newblock Variational principles for stochastic fluid dynamics.
\newblock {\em Proc. R. Soc. A: Math. Phys. Eng. Sci.}, 471(2176):20140963, 2015.

\bibitem{holm1998euler}
D.~D. Holm, J.~E. Marsden, and T.~S. Ratiu.
\newblock Euler-poincar{\'e} models of ideal fluids with nonlinear dispersion.
\newblock {\em Phys. Rev. Lett.}, 80(19):4173, 1998.

\bibitem{hoppe1989diffeomorphism}
J.~Hoppe.
\newblock Diffeomorphism groups, quantization, and {SU}($\infty$).
\newblock {\em Internat. J. Modern Phys. A}, 4(19):5235--5248, 1989.

\bibitem{hoppe1998some}
J.~Hoppe and S.-T. Yau.
\newblock Some properties of matrix harmonics on s 2.
\newblock {\em Comm. Math. Phys.}, 195:67--77, 1998.

\bibitem{kraichnan1967inertial}
R.~H. Kraichnan.
\newblock Inertial ranges in two-dimensional turbulence.
\newblock {\em Phys. Fluids}, 10(7):1417--1423, 1967.

\bibitem{leray1934mouvement}
J.~Leray.
\newblock Sur le mouvement d'un liquide visqueux emplissant l'espace.
\newblock {\em Acta Math.}, 63:193--248, 1934.

\bibitem{lindborg2022two}
E.~Lindborg and A.~Nordmark.
\newblock Two-dimensional turbulence on a sphere.
\newblock {\em J. Fluid Mech.}, 933:A60, 2022.

\bibitem{luesink2024geometric}
E.~Luesink, A.~Franken, S.~Ephrati, and B.~Geurts.
\newblock Geometric derivation and structure-preserving simulation of quasi-geostrophy on the sphere.
\newblock {\em arXiv preprint arXiv:2402.13707}, 2024.

\bibitem{lunasin2007study}
E.~Lunasin, S.~Kurien, M.~A. Taylor, and E.~S. Titi.
\newblock A study of the {N}avier–{S}tokes-$\alpha$ model for two-dimensional turbulence.
\newblock {\em J. Turbul.}, 8:N30, 2007.

\bibitem{lunasin2008spectral}
E.~Lunasin, S.~Kurien, and E.~S. Titi.
\newblock Spectral scaling of the leray-$\alpha$ model for two-dimensional turbulence.
\newblock {\em J. Phys. A}, 41(34):344014, 2008.

\bibitem{marsden2013introduction}
J.~E. Marsden and T.~S. Ratiu.
\newblock {\em Introduction to mechanics and symmetry: a basic exposition of classical mechanical systems}, volume~17.
\newblock Springer, 2013.

\bibitem{marsden2003anisotropic}
J.~E. Marsden and S.~Shkoller.
\newblock The anisotropic lagrangian averaged euler and navier-stokes equations.
\newblock {\em Arch. Ration. Mech. Anal.}, 166:27--46, 2003.

\bibitem{modin2020casimir}
K.~Modin and M.~Viviani.
\newblock A {C}asimir preserving scheme for long-time simulation of spherical ideal hydrodynamics.
\newblock {\em J. Fluid Mech.}, 884:A22, 2020.

\bibitem{modin2020lie}
K.~Modin and M.~Viviani.
\newblock Lie--poisson methods for isospectral flows.
\newblock {\em Found. Comput. Math.}, 20(4):889--921, 2020.

\bibitem{strang1968construction}
G.~Strang.
\newblock On the construction and comparison of difference schemes.
\newblock {\em SIAM J. Numer. Anal.}, 5(3):506--517, 1968.

\bibitem{zeitlin1991finite}
V.~Zeitlin.
\newblock Finite-mode analogs of 2d ideal hydrodynamics: Coadjoint orbits and local canonical structure.
\newblock {\em Phys. D}, 49(3):353--362, 1991.

\bibitem{zeitlin2004self}
V.~Zeitlin.
\newblock Self-consistent finite-mode approximations for the hydrodynamics of an incompressible fluid on nonrotating and rotating spheres.
\newblock {\em Phys. Rev. Lett.}, 93(26):264501, 2004.

\end{thebibliography}

\end{document}